# Songlines and Navigation in Wardaman and other Australian Aboriginal Cultures

Ray P. Norris[1,2] and Bill Yidumduma Harney[3]

[1] CSIRO Astronomy and Space Science, PO Box 76, Epping, NSW, 1710, Australia
[2] Warawara - Department of Indigenous Studies, Macquarie University, NSW, 2109, Australia
[3] Senior Wardaman Elder, PO Box 1579, Katherine, NT, 0851, Australia
*Corresponding Email: Ray.Norris@csiro.au*

**Abstract:** We discuss the songlines and navigation of the Wardaman people, and place them in context by comparing them with corresponding practices in other Australian Aboriginal language groups, using previously unpublished information and also information drawn from the literature. Songlines are effectively oral maps of the landscape, enabling the transmission of oral navigational skills in cultures that do not have a written language. In many cases, songlines on the earth are mirrored by songlines in the sky, enabling the sky to be used as a navigational tool, both by using it as a compass, and by using it as a mnemonic

**Notice to Aboriginal and Torres Strait Islander Readers**: This paper contains the names of people who have passed away.

## 1. INTRODUCTION

### 1.1 AUSTRALIAN ABORIGINAL ASTRONOMY

It is now well established that many traditional Aboriginal cultures incorporate significant references to the sky and to astronomical phenomena (e.g. Stanbridge, 1857, 1861; Mountford, 1956, 1976; Haynes, 1992; Johnson, 1998; Cairns and Harney, 2004; Norris and Norris, 2009; Norris and Hamacher, 2009, 2011; Fuller et al., 2014a). For example, many different Aboriginal cultures across Australia refer to the "Emu in the Sky" (Massola, 1963; Cairns and Harney, 2004; Norris and Norris, 2009, Fuller et al., 2014b), formed from the arrangement of dark clouds within the Milky Way. Equally important in many Aboriginal cultures across Australia are the Orion constellation, which usually symbolises a young man or group of young men, and the Pleiades (Seven Sisters) cluster, which usually symbolises a group of girls pursued by Orion. Star stories can also encapsulate ceremony, law, and culture for transmission to the next generation (Harney and Norris, 2009).

This traditional knowledge extends well beyond mere symbolism, and many Aboriginal cultures contain evidence of a detailed understanding of the sky. For example, within traditional songs can be found explanations of tides, eclipses, and the motion of the celestial bodies (Norris, 2007; Norris and Norris, 2009; Hamacher and Norris, 2012). Practical applications of this knowledge include the ability to predict tides, as well as navigation, time keeping, and the maintenance of a calendar (Cairns and Harney, 2004; Clarke, 2009).

Evidence for these astronomical traditions are found not only in oral traditions, but also in art and artifacts. Some groups of stone arrangements are aligned to cardinal points with an accuracy attainable only by astronomical measurement (Hamacher et al., 2013). The Wurdi Youang stone ring in Victoria contains alignments to the position of the setting sun at the equinox and the solstices (Norris et al., 2013). Statistical tests show that these alignments are unlikely to have arisen by chance, and instead the builders of this stone arrangement appear to have deliberately aligned the site to astronomically significant positions.

Although evidence of astronomical knowledge has been found in many Aboriginal cultures, the best-documented example is undoubtedly that of the Wardaman people, largely because of co-author Harney's enthusiasm to share his traditional knowledge with the wider world. In particular, the book Dark Sparklers (Cairns and Harney, 2004: henceforth DS) documents in exquisite detail the astronomical lore of the Wardaman people.

## 1.2      DIRECTIONALITY

The concept of cardinal directions is common amongst Aboriginal language groups in Australia (Hamacher et al., 2013, and references therein). The Warlpiri people in central Australia are especially prominent in this respect, as much of their culture is based on the four cardinal directions that correspond closely to the four cardinal points (north, south, east, west) of modern western culture (Laughren, 1978, 1992; Nash, 1980, Pawu-Kurlpurlurnu et al., 2008). In the Warlpiri culture, north corresponds to "law", south to "ceremony", west to "language", and east to "skin". "Country" lies at the intersection of these directions, at the centre of the compass  - i.e. "here" (Pawu-Kurlpurlurnu, 2008; also personal communication from Pawu-Kurlpurlurnu to Norris, 2008).

Cardinal directions are also important in Wardaman culture, and were created in the Dreaming by the Blue-tongued Lizard (DS: 60):

> "*Blue-tongue Lungarra now he showing all these boomerang, calling out*
> *all the names: east, west, north, south, all these sort of type.*"

Other language groups have cardinal directions that may vary from the modern western convention, although east and west are often associated with the rising and setting positions of the sun, and the words for east and west are often based on the word of the sun (Hamacher et al., 2013, and references therein). However, in some cases, the cardinal positions are loosely defined and may vary markedly from place to place (Breen, 1993).

Directionality is also important in the sleeping position, as described by Harney (DS: 61):

> *"We gotta sleep east, not downhill. We can sleep crossway, but we're not allowed to sleep towards the sun going down. Sleep down the bottom, its bad luck for you because you're against the sun. If you sleep on the eastern way and going that away, that's fine. Facing west, you gotta change your bed. Head up on the east when you sleep … each person where they die, in our Law, we always face them to their country. Graveyard always face to their country, they can look straight to their country."*

Burials in other traditional Aboriginal cultures were often aligned to cardinal directions. For example, the deceased in NSW (New South Wales) were buried facing east, in a sitting position (Dunbar, 1943; Mathews, 1904: 274).

In contrast to the east-west alignments of burials, initiation sites in NSW were often aligned roughly north-south. In a study of Bora (initiation) sites in NSW, Fuller et al. (2013) found that bora sites have a statistically significant preferred orientation to the south and southwest, which is consistent with circumstantial evidence that bora grounds are aligned with the position of the Milky Way in the night sky in August, which is roughly vertical in the evening sky to the south-southwest. This connection between Bora sites and the Milky Way was subsequently confirmed in ethnographic studies by Fuller et al. (2014b), who found that the head and body of the Emu in the Sky correspond to the small and large Bora rings respectively on the ground.

## 1.3    WARDAMAN ASTRONOMY

The land of the Wardaman people is about 200 km southwest of Katherine in the Northern Territory (see Figure 1). We are fortunate that co-author Harney grew up and was educated and initiated at a time when the Wardaman people still followed a largely traditional lifestyle (see Figures 2 and 3). The language and culture of the Wardaman people has a particularly strong astronomical component, and are well documented in DS. The three major creation figures (Froglady Earthmother and her two husbands, Rainbow and Sky Boss) are all signified by dark clouds in the Milky Way, and stars and nebulae document other figures and other events. The Southern Cross is particularly important, and its orientation defines the Wardaman calendar and marks the cycle of dreaming stories throughout the year.

In the words of Harney (2009):

> *"In the country the landscape, the walking and dark on foot all around the country in the long grass, spearing, hunting, gathering with our Mum and all this but each night where we were going to travel back to the camp otherwise you don't get lost and all the only tell was about a star. How to travel? Follow the star along. … While we were growing up. We only lay on our back and talk about the stars. We talk about emus and kangaroos, the whole and the stars, the turkeys and the willy wagtail, the whole lot, everything up in the star we named them all with Aboriginal* *names. Anyway we talked about a lot of that ... but we didn't have a watch in those days. We always followed the star for the watch. ... Emu, Crocodile, Cat Fish, Eagle Hawk, and all in the sky in one of the stars. The stars and the Milky Way have been moving all around. If you lay on your back in the middle of the night you can see the stars all blinking. They're all talking."*

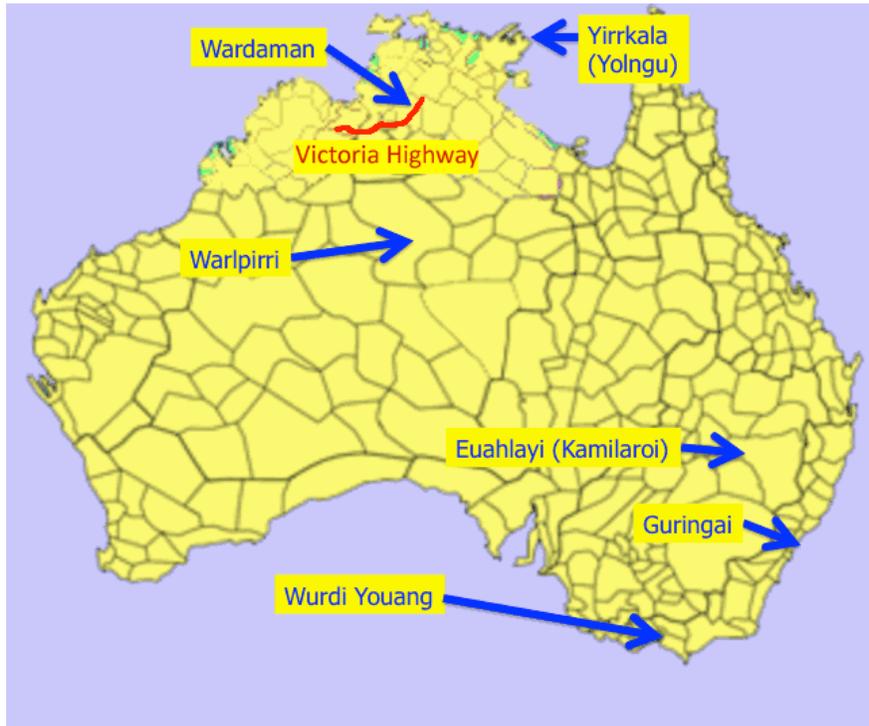

*Figure 1: Map showing some of the places and the locations of language groups discussed in the text. Adapted from a map licensed under the* <u>Creative Commons</u> *Attribution-Share Alike 3.0*

## 1.4    This Paper

Aboriginal songlines and navigation are not well documented, and the primary goal of this paper is to summarise the available information, including some previously unpublished information. This paper is doubtless incomplete, and will hopefully be supplemented or superseded by more detailed studies.

This paper focuses on the songlines and navigation of the Wardaman people, for which the best documentation is available, while making comparisons with corresponding practices in other Australian Aboriginal language groups. This paper departs from conventional scholarly practice because one of the authors (Harney) is the senior Wardaman elder with a great reserve of traditional knowledge, much of which has not yet been documented. It is therefore appropriate to include quotes from Harney in his own words. One aim of this paper is to document some of this traditional knowledge. Where possible, we do so by using the verbatim transcripts of Harney's verbal descriptions, shown in italics and accompanied by a reference with dates and other details. No attempt has been made to reword these descriptions. This is to retain the original flavour and avoid unintentional misinterpretation.

We recognise that the many different Aboriginal language groups have different practices and cultures, and by describing the practices of different cultures in this paper we recognise the risk of imposing a "one size fits all" stereotype to all these cultures. That is not our intention. Instead, this paper should be regarded as a sample of the available information on the navigational practices of Aboriginal Australians.

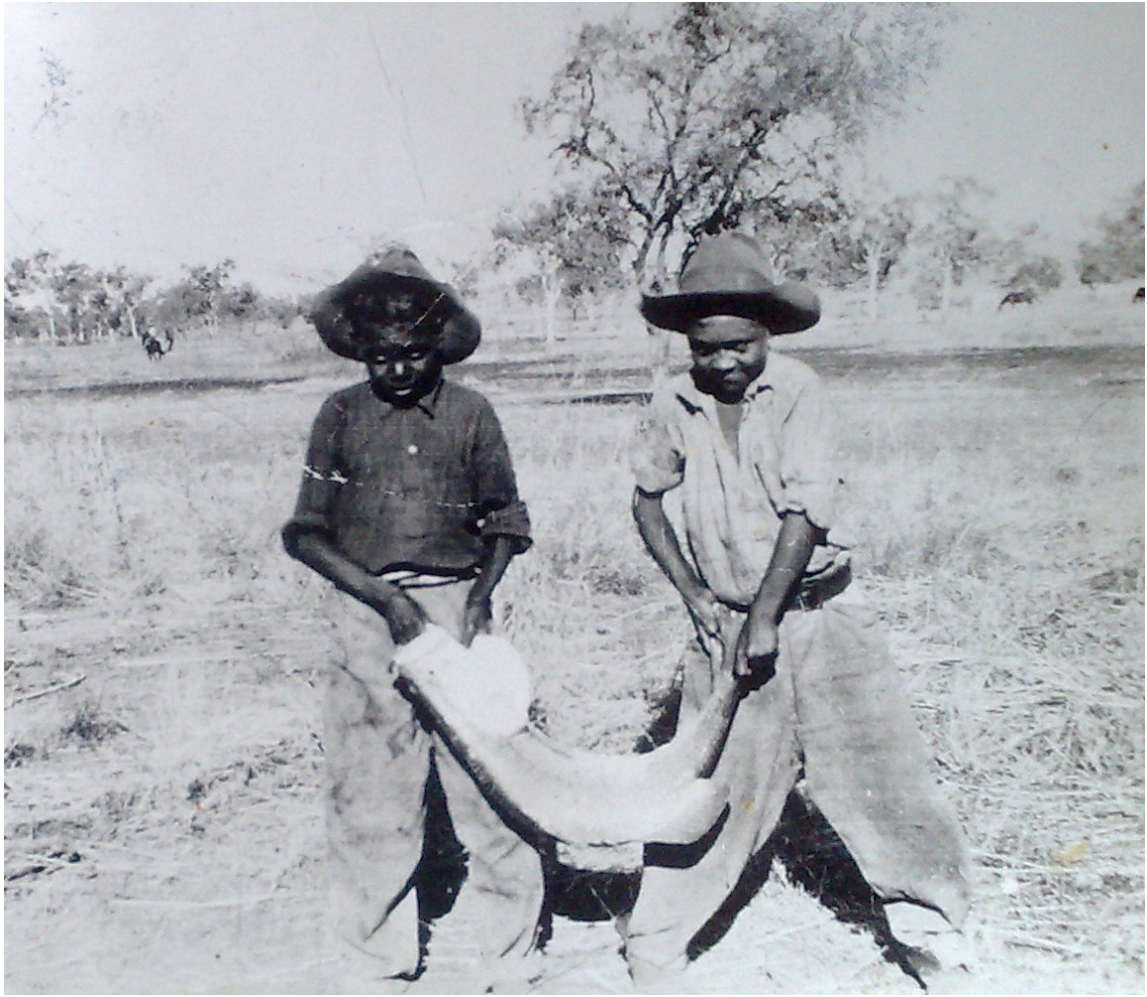

*Figure 2: Previously unpublished photo of Bill Yidumduma Harney (right) in about 1940, together with his friend and a large barramundi they had caught in a waterhole on Willeroo Station. Photo courtesy of B.Y. Harney.*

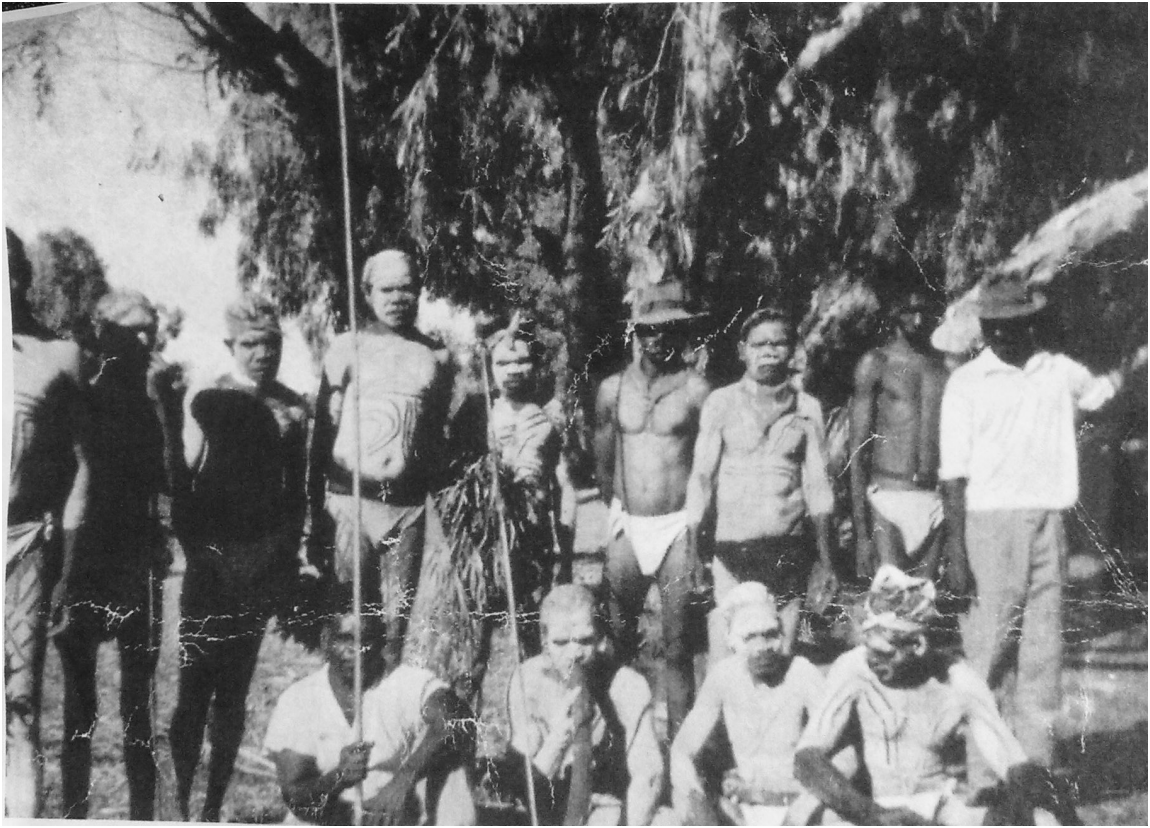

*Figure 3: Previously unpublished photo of Bill Yidumduma Harney's family in 1929. Photo courtesy of B.Y. Harney.*

## 2    SONGLINES

The English word "songline" was coined by Chatwin (1987), but the concept is ancient and embedded in traditional Aboriginal cultures. They are often referred to as "Dreaming Tracks", and can also be called "strings" (Clarke, 2003: 19) in the sense that they connect different people and sacred sites. According to Harney (2009):

> "*Between the star and the landscape and the rock painting and all that, they're more or less connected together around the country.*"

Mulvaney and Kamminga (1999: 95) argue that they also represent the trading routes that criss-cross Australia. Gammage (2011: 24) says:

> "*A songline or storyline is the path or corridor along which a creator ancestor moved to bring country into being. It is also the way of the ancestor's totem, the geographical expression of their songs, dances and paintings animating its country, and ecological proof of the unity of things.*"

According to Wositsky and Harney (1999: 301):

> *"Songlines are epic creation songs passed to present generations by a line of singers continuous since the dreamtime. These songs, or song-cycles, have various names according to which language group they belong to, and tell the story of the creation of the land, provide maps for the country, and hand down law as decreed by the creation heroes of the dreamtime. Some songlines describe a path crossing the entire Australian continent."*

As well as marking routes on the ground, songlines were also paths in the sky, several examples of which are described in detail in DS. Harney (2009, personal communication to Norris) described how the songlines on Earth were mirrored by the songlines in the sky, so that knowledge of the sky formed a mnemonic for tracing a route on Earth. This mirroring was created when the Creator Spirits moved to the sky (DS: 99):

> *"One day it was all different, when they come down and make up the Creation line songs, because they travelling. When everything become still. They all split up, land, become all the stars..."*

For example, one songline starts at Yirrkala in Arnhem Land, where the Yolngu believe Barnumbirr (Venus) crossed the coast as she brought the first humans to Australia from the east (Allen, 1975; Norris and Norris, 2009). Her song, contained within the Yolngu Morning Star ceremony, describes her path across the land, including the location of mountains, waterholes, landmarks, and boundaries. The song therefore constitutes an oral map, enabling the traveller to navigate across the land while finding food and water. It is said by Yolngu elders at Yirrkala that the same song is recognisable in a number of different languages along the path from east to west, crossing the entire "top end" of Australia. The song changes along the route, being longer and more "sing-song" in the east, and shorter, and broken into short sharp segments, in the west (personal communication by elders at Yirrkala to Norris, 2007).

Many other songlines are known across Australia (e.g. Kerwin, 2010). Fuller et al. (2014c) report several songlines known by the Euahlayi people, including the eaglehawk songline that extends from Heavitree Gap at Alice Springs to Byron Bay on the East Coast, connecting the Arrernte people with the Euahlayi people, and also connecting the stars Achernar, Canopus, and Sirius. The Euahlayi people also know the Black Snake/Bogong Moth songline connecting Normanton on the Gulf of Carpentaria with the Snowy Mountains near Canberra, and which also follows the Milky Way.

Another example is the two songlines that are said by Darug elders to extend west from Sydney, through Sackville, and then roughly following the paths of the Great Western Highway and the Bells Line of Road respectively, until they join again at Little Hartley (personal communication by Des Dyer and Gordon Workman of the Darug Tribal Aboriginal Corporation to Norris, 2007). Supporting evidence includes the Darug rock engravings found close to the path of the Great Western Highway through the Blue Mountains.

The creation of the songlines is described by Harney and Lee (2010: 11):

> *"They put all them together, then with that, they made all the Songlines, right across the country. And that Creation Song now, we still got it today. Nothing been changed, we still got that old one. Original one. Because we gotta have that for all this rock painting, all the different sites, and rock."*

Such long distance paths were important because of the important trading routes (e.g. Gammage, 2011; Kerwin, 2010, and references therein) traversing Australia for the trading and exchange of goods, such as the export of ochre from Wardaman country. Lee and Harney (2009) explain that:

> "*Red and yellow from this area are considered very powerful and were traded for long distances for use in ceremonies.*"

A journey, sometimes taking months, would have several functions, including attending ceremonies, as well as trading. The trading itself might include the trading of intellectual property, such as songs and dances, as well as material objects.

Later, many of these ancient trade routes, many based on songlines, laid the basis for some of the current network of highways across Australia (Wositsky and Harney, 1999: 14):

> *"They showed him the way right through from Willeroo to Victoria River Downs Station. My grandfather used to go in the lead, and blaze the trees all the way, following the old Aboriginal walking pad, and old Bill would come along behind him making the road. They call it the Victoria Highway now, but it was never the Victoria Highway at all – it was just the original Aboriginal walking trail right through Arnhem Land and Katherine Gorge and past Willeroo, right down to Western Australia. They used that walking trail to trade their boomerangs and spears and many different ochres and when they did the trade they had ceremonial meetings. That had been a walking trail for a hell of a long time… all the way from Cape York right through Borroloola and straight across the country. They hit Roper River and followed the Roper River all the way past Mataranka and they came right past Willeroo."*

## 3 NAVIGATION

Tindale (1974: 75) was aware of ancient Aboriginal tracks across large parts of Australia, but considered that their use was mainly for travelling short distances. However he noted of the trading routes (*ibid*: 81) that:

> "*the great distances covered and also the difficulties encountered, considering the precarious line of communication across formidable dry areas, are striking.*"

It is curious that there is almost no discussion of the navigational practices of Aboriginal people by those who studied their culture extensively in the last century, such as Elkin, the Berndts, and Mountford. With hindsight, many of the songs and stories that they describe involve a route on earth, or in the sky, followed by creator-spirits, but were not discussed at the time in terms of navigation. For example, Elkin (1938: 304) appears to have heard at least one songline without noting its significance:

> "*each… sings all night its cycle of the hero's experiences as he journeyed from the north coast south and then back again north… a headman sitting nearby commented that the Ngurlmak, according to the text, was now in that country, then in another place, and so on, ever coming nearer until at last it was just where we were making the recording.*"

Mountford (1976: 50) discussed the extensive trading networks without asking how people navigated these vast distances, and discussed Aboriginal astronomy without asking if the stars were used for navigation. He apparently encountered song-line descriptions, but didn't remark on their navigational significance. For example, he recounts (*ibid*: 462):

> "*The series of twelve drawings… indicate that the route of Orion and the Pleiades extends from the Warburton range in Western Australia through the Rawlison, Petermann, Mann, and Musgrave ranges, reaching Glen Helen, in the country of the Western Aranda. At some point between the Petermann and Mann mythical route, the name of the man of Orion was changed from Jula to Nirunja […].*"

It is unclear whether this lack of discussion reflects the assumptions and interests of the anthropologists at the time, or because this knowledge was regarded by the Aboriginal participants as secret. Nevertheless, the available evidence (e.g. Kerwin, 2010) shows unambiguously that the ability to navigate long distances was widespread.

Amongst the Wardaman people in northern Australia, most travelling was done at night, when the air was cool and the stars visible as guides. Furthermore, there was a belief that distances were smaller at night (DS: 65):

> "*The old people, the old man walking during the day saying the distance get far away from you. Walk in the night in the darkness, the distance shrinking up. Somehow it's shrinking up! The earth's pulling away from you pretty fast! Shrinking up, that's what they told us. But … during the daytime, the earth's still standing still. Aborigine call it in a word, the Shame – he shamed to move.*"

Elsewhere in Australia, nighttime travel was less common. Maegraith (1932) and Lewis (1976) found that Central Desert people did not use the stars for navigation, and did not travel at night, and Fuller et al. (2014c) found that Euahlayi people did not travel at night.

Where possible, and for long-distance navigation, a songline would be followed (DS: 63):

> "*Not just songline trail, walking trail, trade routes. You sing a song, then you follow your song, in that track you go along singing the song, like a blazed mark.*"

Traditional Aboriginal elders (such as Harney) have an intimate knowledge of the night sky, far better than that of most modern-day western astronomers (such as Norris), and can name many stars in any given patch of sky, and explain their role in the Dreaming stories. Mountford (1976: 449) considered that:

> "*many Aborigines of the desert are aware of every star in their firmament, down to at least the fourth magnitude, and most, if not all, of those stars would have myths associated with them.*"

These Aboriginal elders also understood how the whole pattern rotated over their heads from east to west during the night, and how it shifted over the course of a year (DS: 61):

> "*Each time you look at the stars it's in a different inch by half an inch, or quarter of an inch by quarter of an inch, or whatever. That's where the old Aborigines, all the elders, see that travelling. Then later on it's over there earlier in the year.*"

For shorter journeys, or when a songline was not available, the direction of the Moon or patterns of the stars were used for navigation (DS: 63):

> "*You judge how far it is to Willeroo, you say about 3km, you aim for that 3 km in your mind. That's all! You've got to go cross ways? That's Emu Foot tells you, he's south. If you want to go southwest, you go on the right hand side of the emu […]*"

Navigating by the stars is still considered by Harney (2009) to be preferable to following the modern road:

> "*You know road might be going to the water a bit, road might be going out of the waterhole and you like to get perished too that's what he's about. But the star the mainly the one really guide you straight to the waterhole and all this.*"

The path of the planets in the sky, the ecliptic, had special significance (DS: 65):

> *"The Dreaming Track in the sky! Planets making the pathway! Travelling routes, a pathway you could call it, like a highway! Travelling pathway joins to all different areas, to base place, to camping place, to ceremony place, where the trade routes come in; all this sort of things. The Dreaming Track in the sky, the planets come straight across… walking trail becomes a pad, then becomes a wagon road, two wheel tracks, then become a highway. That's how they started off, four of them."*

While star maps do exist in Aboriginal paintings and possibly in rock engravings, no Aboriginal star maps intended for navigation have been recorded. Instead, all the knowledge is committed to memory in the form of songlines, which may therefore be regarded as "oral maps". In a culture with no written language, but with a strong tradition of memorising oral knowledge, this is probably the optimum way of recording and transmitting navigational information.

## 4    CONCLUDING REMARKS

We have presented new information, and also material drawn from the literature, to show that:

1. Songlines are effectively oral maps of the landscape, enabling the transmission of oral navigational skills in cultures that do not have a written language,
2. Songlines extend for large distances across Australia, and are often identical to the trading routes, and were presumably used for navigating these trading routes,
3. Some modern Australian highways follow the path of Aboriginal songlines,
4. In many cases, songlines on the earth are mirrored by songlines in the sky,
5. The sky is used as a navigational tool, both by using it as a compass, and by using it as a mnemonic to remember the songlines on the ground.

## 5    ACKNOWLEDGEMENT

We acknowledge and pay our respects to the traditional owners and elders, both past and present, of the Wardaman people and of all the other language groups mentioned in this paper. We also thank Hugh Cairns for permission to reproduce selected passages from *Dark Sparklers*.

## 6    REFERENCES

Allen, L.A., 1975. *Time before morning*. New York, Cowell.

Breen, G., 1993. *East is south and west is north. Australian Aboriginal Studies,* 1993(2), 20–34.

Cairns, H., and Harney, B.Y., 2004. *Dark sparklers Yidumduma's Wardaman Aboriginal astronomy (revised edition)*. Merimbula, NSW, H.C. Cairns.

Chatwin, B., 1987. *The Songlines*. London, Jonathan Cape.

Clarke, P.A., 2003. *Where the ancestors walked: Australia as an Aboriginal landscape*. Sydney, Allen and Unwin.

Clarke, P.A., 2009. Australian Aboriginal ethnometeorology and seasonal calendars. *History and Anthropology*, 20, 79-106.

Dunbar, G.K., 1943. Notes on the Ngemba tribe of the central Darling River, western NSW. *Mankind*, 3(5), 140–148.

Elkin, A.P., 1938. *The Australian Aborigines: How to Understand Them*. Sydney, Angus and Robertson.

Fuller, R.S., Hamacher, D.W., and Norris, R.P., 2013. Astronomical orientations of Bora ceremonial grounds in southeast Australia. *Australian Archaeology*, 77, 30-37.

Fuller, R.S., Norris, R.P., and Trudgett, M., 2014a. The astronomy of the Kamilaroi people and their neighbou*r*s. *Australian Aboriginal Studies*, in press. Preprint: http://arXiv.org/abs/1311.0076

Fuller, R.S., Anderson, M.G., Norris, R.P., Trudgett, M., 2014b. The emu sky knowledge of the Kamilaroi and Euahlayi peoples. *Journal of Astronomical History and Heritage*, 17(2), this volume.

Fuller, R.S, Anderson, M.G., Norris, R.P., Trudgett, M., 2014c. Star maps and travelling to ceremonies: the Euahlayi people and their use of the night sky. *Journal of Astronomical History and Heritage*, in review.

Gammage, B., 2011. *The biggest estate on Earth*. Sydney, Allen and Unwin.

Hamacher, D.W. and Norris, R.P., 2012. Eclipses in Australian Aboriginal astronomy. *Journal of Astronomical History and Heritage*, 14(2), 103-114

Hamacher, D.W., Fuller, R.S., and Norris, R.P., 2013. Orientations of linear stone arrangements in New South Wales. *Australian Archaeology,* 75, 46-54.

Harney, B.Y., 2009. *Wardaman astronomy*. Unpublished presentation at "Ilgarijiri - things belonging to the sky: a symposium on Australian Indigenous Astronomy" held on 27 November 2009 at the Australian Institute for Aboriginal and Torres Strait Islander Studies (AIATSIS) in Canberra, ACT, Australia.

Harney, B.Y., and Norris, R.P., 2009. Unpublished presentation *The First Astronomers?* Darwin Festival, August 2009, Darwin, NT, Australia. View excerpts here: http://www.youtube.com/watch?v=TKMnaZQp418

Harney, B.Y., and Lee, D.M., 2010. *Wardaman creation story*. Bishop, CA, D.M. Lee.

Haynes, R.D., 1992. Aboriginal astronomy. *Australian Journal of Astronomy*, 4, 127-140.

Johnson, D.D., 1998. *Night skies of Aboriginal Australia: a noctuary*. Oceania Monograph No. 47. Sydney, University of Sydney Press.

Kerwin, D., 2010. *Aboriginal Dreaming paths and trading routes: the colonisation of the Australian economic landscape*. Eastbourne, UK, Sussex Academic Press.

Laughren, M.N., 1978. *Directional terminology in Warlpiri (a Central Australian language)*. Working Papers in Language and Linguistics, Volume. 8. Launceston, Tasmanian College of Advanced Education. Pp. 1-16.

Laughren, M.N., 1992. Cardinal directions in Warlpiri. Unpublished paper presented to the Space in Language and Interaction in Aboriginal Australia Workshop, Australian Linguistic Institute, Sydney.

Lee, D.M., and Harney, B.Y., 2009. *Introduction to the rock art of Wardaman country*. Bishop, CA, D.M. Lee.

Lewis, D., 1976. *Observations on route finding and spatial orientation among the Aboriginal peoples of the Western Desert region of Central Australia*. Oceania, 46(4), 249-282.

Maegraith, B., 1932. The astronomy of the Aranda and Luritja tribes. *Transactions of the Royal Society of South Australia*, 56 (1), 19-26.

Massola, A., 1963. Native stone arrangement at Carisbrook. *The Victorian Naturalist*, 80, 177-180.

Mathews, R.H., 1904. Ethnographical notes on the Aboriginal tribes of New South Wales and Victoria. *Journal and Proceedings of the Royal Society of New South Wales*, 38, 203–381.

Mountford, C.P., 1956. *Records of the American-Australian Scientific Expedition to Arnhem Land, Volume 1*. Melbourne, University of Melbourne.

Mountford, C.P., 1976. *Nomads of the Australian desert*. Adelaide, Rigby.

Mulvaney, D.J., and Kamminga, J., 1999. *Prehistory of Australia*. Washington DC, Smithsonian Institution Press.

Nash, D.G., 1980. *Topics in Warlpiri grammar.* Unpublished PhD Thesis, Department of Linguistics and Philosophy, Massachusetts Institute of Technology. Cambridge, MA, USA.

Norris, R.P., 2007. *Searching for the astronomy of Aboriginal Australians.* In J. Vaiskunas (edt) *Astronomy and Cosmology in Folk Traditions and Cultural Heritage.* Archaeologia Baltica, Volume 10. Klaipėda: Klaipėda University Press. Pp. 246-252.

Norris, R.P., and Norris, C.M., 2009. *Emu Dreaming: an introduction to Australian Aboriginal astronomy.* Sydney, Emu Dreaming Press.

Norris, R.P., and Hamacher D.W., 2009. *The astronomy of Aboriginal Australia.* In D. Valls-Gabaud and A. Boksenberg (eds) *The role of Astronomy in Society and Culture.* Cambridge, Cambridge University Press. Pp. 39-47

Norris, R.P., and Hamacher, D.W., 2011. Astronomical symbolism in Australian Aboriginal rock art. *Rock Art Research,* 28, 99-106.

Norris, R.P, Norris, P.M., Hamacher, D.W., and Abrahams, R., 2013. Wurdi Youang: an Australian Aboriginal stone arrangement with possible solar indications. *Rock Art Research*, 30(1), 55-65.

Pawu-Kurlpurlurnu, W.J., Holmes, M., and Box, L., 2008. *Ngurra-kurlu: a way of working with Warlpiri people.* DKCRC Report 41. Alice Springs, NT, Desert Knowledge Cooperative Research Centre.

Stanbridge, W.E., 1857. On the astronomy and mythology of the Aborigines of Victoria. *Transactions of the Philosophical Institute of Victoria*, 2, 137-140.

Stanbridge, W.E., 1861. Some particulars of the general characteristics, astronomy, and mythology of the tribes in the central part of Victoria, southern Australia. *Transactions of the Ethnological Society of London*, 1, 286-304.

Tindale, N.B., 1974. *Aboriginal tribes of Australia.* Berkeley, University of California Press.

Wositsky, J., and Harney, B.Y., 1999. *Born under the paperbark tree.* Marlston, SA, JB Books.

## ABOUT THE AUTHORS

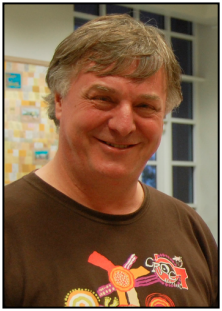

Professor Ray Norris is an astrophysicist with CSIRO Astronomy & Space Science and an Adjunct Professor at Warawara - Department of Indigenous Studies at Macquarie University, both in Sydney, Australia. He has published about 300 peer-reviewed papers, including 15 on Aboriginal Astronomy, and wrote the book *Emu Dreaming: an Introduction to Australian Aboriginal Astronomy*.

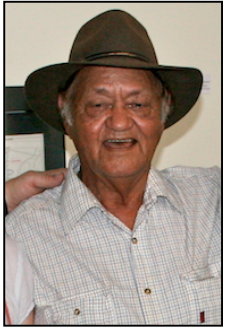

Bill Yidumduma Harney is an Elder and the last fully initiated Senior Custodian of the Wardaman people near Katherine, NT, Australia. He is an esteemed artist, master storyteller, and musician. Well known as an advocate and ambassador for Aboriginal Australians, Yidumduma was raised and educated in the traditional ceremonies by Jumorji, a senior Wardaman lawman and his Aboriginal stepfather. He is of the Yubulyawan clan, speaks seven languages, and co-authored the book *Dark Sparklers* (2004) about Wardaman astronomy with Hugh Cairns.